\documentstyle[11pt]{article}
\begin{document}\openup6pt
\catcode`\@=11

\title{EMERGENT UNIVERSE IN STAROBINSKY MODEL}

\author{S. Mukherjee\thanks{Email: sailom@iucaa.ernet.in}     
 and B.C. Paul\thanks{Email: %%@
bcpaul@iucaa.ernet.in }
  \\
Physics Department, North Bengal University \\
Dist : Darjeeling, PIN : 734 430, India. \\
S. D. Maharaj\thanks{E-mail: maharaj@ukzn.ac.za}\\
 Astrophysics and Cosmology Research Unit  \\
School of  Mathematical Sciences, University of KwaZulu-Natal \\
Durban 4041, South Africa \\
and \\
A. Beesham\thanks{E-mail: abeesham@pan.uzulu.ac.za} \\
 Department of Mathematical Sciences, Zululand University \\
Kwadlangezwa, South Africa
}

\date{}
\maketitle
\vspace{0.5in}

\begin{abstract}

 We present an emergent universe scenario  making use of a new solution of  the Starobinsky model.  The solution belongs to a one parameter family of solutions, where the parameter is determined by the number and the species (spin-values) of primordial fields. The general features of the model have also been studied.
  
\end{abstract}

\vspace{0.2cm}
\pagebreak

\section{ INTRODUCTION }

Recently Ellis and Maartens [1] reconsidered the possibility of a cosmological model in 
which there is no big-bang singularity, no beginning of time and the universe 
effectively avoids a quantum regime for space-time by staying large at all times. The 
universe starts out in the infinite past as an almost static universe and expands 
slowly, eventually evolving into a hot big-bang era. An interesting example of this 
scenario is given by Ellis, Murugan and Tsagas [2], for a closed universe model with a 
minimally coupled scalar field $\phi$, which has a special form of  interaction potential $ V 
(\phi)$. It was pointed out that this potential is similar to what one obtains from a $R 
+ \alpha R^{2}$ theory after a suitable conformal transformation   and identifying  $ 
\phi = - \sqrt{3} \; \ln (1 + 2 \alpha R) $ with a negative $\alpha$. Although the probability of these solutions 
is not high, the emergent universe scenario nevertheless merits attention as it solves 
many conceptual and technical problems of the big bang model. In this paper, we point 
out that the Starobinsky model, the original as well as the modified version, permit 
solutions describing an emergent universe. The solution may be used to model varieties
of cosmological scenarios consistent with  the observational results, available at present  
and expected in the near future. Thus, it may be possible to build models which avoid the 
quantum regime for space-time  but share the good features of the standard big bang
model.

In the models considered in $\it Ref.$ [1] $\it and $ [2], a closed universe was considered. 
However,  recent results from BOOMERANG and WMAP indicate that the universe is most likely to be  spatially flat. If the universe has  always
been large enough, the field equations become simpler. In fact, in the Starobinsky model, the field equations can be written as a second 
order differential equation for the Hubble parameter $H$, vide equation (7).

\section{ STAROBINSKY MODEL :}

In the Starobinsky model, one considers the semi-classical Einstein equation,
\begin{equation}
R_{\mu \nu} - \frac{1}{2} g_{\mu \nu} R = - \; 8 \pi G < T_{\mu \nu} >
\end{equation}
where    $< T_{\mu \nu} > $ is the vacuum expectation value of the energy momentum tensor. In 
the case of free, massless, conformally invariant fields the vacuum expectation value  
$< T_{\mu \nu} > $ can be written, with the Robertson Walker metric as,
\begin{equation}
<T_{\mu \nu} > \;  =  K_{1} \; ^{(1)}H_{\mu \nu} +  K_{3} \; ^{(3)}H_{\mu \nu}
\end{equation} 
where $K_{1}$ and $K_{3}$ are numbers and
\begin{equation}
^{(1)}H_{\mu \nu}  =  2 R_{; \mu ; \nu} - 2  g_{\mu \nu} R_{; \sigma}^{; \sigma} +  2 R 
R_{ \mu \nu} - \frac{1}{2} g_{\mu \nu} R^{2},
\end{equation} 
\begin{equation}
^{(3)}H_{\mu \nu}  =  2 R_{\mu}^{ \sigma }  R_{\nu \sigma} - \frac{2}{3} R 
R_{ \mu \nu} - \frac{1}{2} g_{\mu \nu} R^{\sigma \tau} R_{\sigma \tau} + \frac{1}{4} 
g_{\mu \nu} R^{2}.
\end{equation} 
Equation (2), written for $< T^{\mu}_{\mu} >$,  gives the well-known trace-anomaly,
indicating that the conformal invariance is broken  by the regularization process. Note 
that  the two tensors $^{(1)} H_{\mu \nu}$ and $^{(3)} H_{\mu \nu}$ have different 
features. The tensor $^{(1)} H_{\mu \nu}$ is identically conserved and can be obtained by 
varying a local action $A \sim {\large \int } \sqrt{- g} \;  R^{2} \; d^{4}x$. However, one of the 
counter terms to be added to the lagrangian to regularise $< T^{\mu}_{\mu} >$ has the 
form $\mu \sqrt{- g} \; R^{2}$, where $\mu$ is a logarithmically  divergent constant, thus
permitting the possibility of an addition of any  finite number to  $\mu$. Hence,  
$K_{1}$ is not determined and can be given any arbitrary value including a negative one. 
The tensor $^{(3)} H_{\mu \nu}$  is conserved only in a conformally flat universe and  it %%@
cannot be obtained by variation of any local action. The constant $K_{3}$ is fully %%@
determined, once the fields are specified, e.g.,
\begin{equation}
K_{3}  =  \frac{1}{1440 \pi^{2}} \left( N_{o} + \frac{11}{2} N_{1/2} + 31 \; %%@
N_{1}\right),
\end{equation} 
where $N_{I}$ gives the number of quantum fields of spin I. The freedom of choosing %%@
$K_{1}$ arbitrarily has led to the modified Starobinsky [3,4] model, where one adds a  counter term $\; \sim \; \sqrt{- g} \;  R^{2}$, and chooses $K_{1} >> K_{3}$. The theory then becomes %%@
essentially a typical $R^{2}$ -theory. However, the earlier work on Starobinsky models were done in the context of a big bang model. The emergent universe model needs a
different scenario and one makes different choices of  $K_{1}$ and $ K_{3}$, as will be 
shown below.

Let us choose $8 \pi G = 1$ and write the evolution equation (1) of the flat FRW 
universe as
\begin{equation}
\frac{\dot{a}^{2}}{a^{2}}
  = K_{3} \;  \frac{\dot{a}^{4}}{a^{4}} - 6 K_{1} \left[  2 \; \frac{ \dot{a}}{2 a^{2}} 
\frac{d \ddot{a}}{dt} - \frac{\ddot{a}^{2}}{a^{2}} + 2 \frac{ \ddot{a}  %%@
\dot{a}^{2}}{a^{3}} -  3 \frac{\dot{a}^{2}}{a^{2}} \right]
\end{equation} 
which can be written in terms of the Hubble parameter
\begin{equation}
H^{2} \left( \frac{1}{ K_{3}} - H^{2} \right) =  -  \frac{ 6 K_{1}}{K_{3}} \left( 2 H %%@
\ddot{H} + 6 H^{2} \dot{H} - \dot{H}^{2} \right).
\end{equation}

Equation (7) will determine the evolution of the universe when the initial %%@
conditions are provided. 

\section{ EMERGENT UNIVERSE :}

We now look for a solution which describes a universe which exists eternally and always %%@
remains large so that a  classical description is possible at all times. The  matter, as %%@
indicated earlier, will be described by quantum field theories. We first note the %%@
following two exact solutions :

$\bullet$ $ H = 0 $ is a solution but it describes an eternally static universe. It can %%@
be checked that the solution  is stable against  small linear perturbations, if $K_{1}$ is %%@
positive. However, with $K_{1}$ negative, the state is unstable.

$\bullet $ 
$H = \frac{1}{\sqrt{K_{3}}} $ gives a de Sitter type solution, but the solution is not %%@
stable.

To get an emergent universe, we look for a solution of the suggestive form
\begin{equation}
a (t) = a_{o} \left( \beta + e^{\alpha t}\right)^{\omega}
\end{equation} 
where the constants $\alpha$ and $\omega$ will be determined from equation (7). The %%@
constants $a_{o}$ and $\beta$ may be determined from  initial conditions. Since we %%@
have
\begin{equation}
H = \frac{\omega \alpha e^{\alpha t}}{\beta + e^{\alpha t}}, \; \dot{H} = 
\frac{\omega \alpha^{2} e^{\alpha t}}{(\beta + e^{\alpha t})^{2}}, \; \ddot{H} = %%@
\frac{\omega \alpha^{3} e^{\alpha t} (1 -  e^{\alpha t})}{(\beta + e^{\alpha t})^{3}}
\end{equation} 
the function in (8)  will be a solution, if
$\omega = \frac{2}{3} , \; \alpha = \frac{3}{2} \sqrt{ \frac{1}{K_{3}}} $ and $ K_{1} = %%@
- \frac{2}{27} K_{3}$. Thus $K_{1}$ will be chosen negative in this model.

The fact that the emergent universe scenario is indeed permitted by equation (7)  comes %%@
as a surprise. Since $K_{1} $ and $K_{3}$  are related, we have a one parameter family %%@
of solutions specified by the value of $K_{3}$. The general features of the solutions %%@
are

1) The scale factor $a(t) $ has a non-zero value $a ( t \rightarrow - \infty) = a_{o} %%@
\beta^{2/3}$ as $t \rightarrow - \infty$, which may be chosen much larger than the %%@
Planck length. The Hubble parameter $H$ and its %%@
time derivatives $\dot{H}$ and $\ddot{H}$ all vanish in the limit $ t \rightarrow - %%@
\infty $. Thus the solution decribes an emergent universe with natural initial %%@
conditions.

2) If $ \beta >> 1$, the universe remains almost static during the period $- \infty < t %%@
< \frac{2}{3} \sqrt{ K_{3}} \; \ln \beta = t_{o}$. Thus during the entire infinite past of  $t = %%@
0$, the universe expands only by a factor 

\begin{equation}
\frac{a( t = 0 )}{a( t \rightarrow - \infty)} = \left( 1 + \frac{1}{\beta} \right)^{2/3} %%@
\sim 1
\end{equation}
for a large $\beta$. 

\vspace{1.0cm}

\input{epsf}
\begin{figure} 
\begin{center}
\vspace{6cm}
\special{eps: 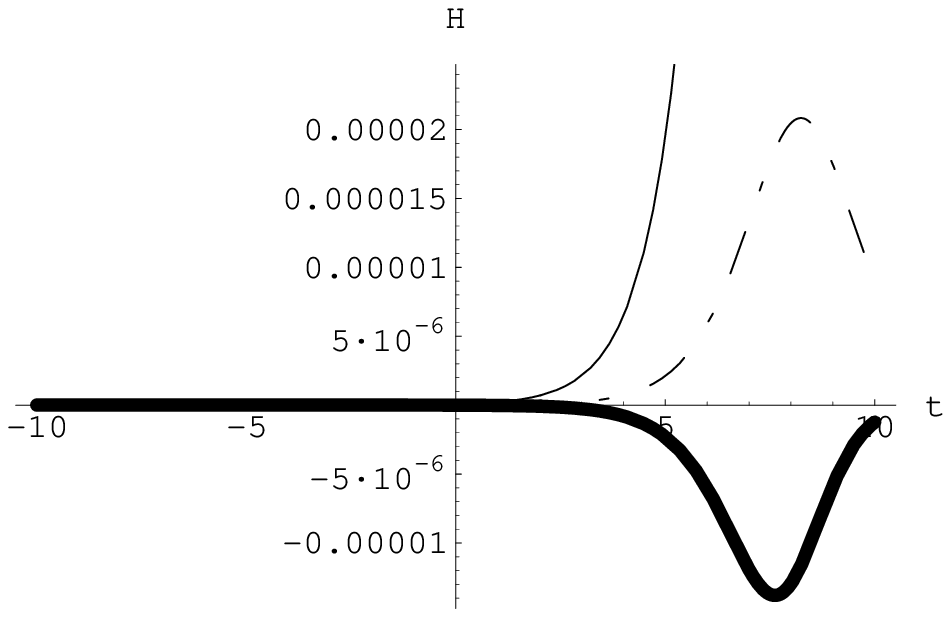 x=10cm y=6cm}
\caption{Variation of $h = H X 10^{-2}$ and the derivatives, $\dot{H}$ and $\ddot{H}$ are 
represented by  thin, broken and  thick lines respectively ( $\beta = 10000 $ and $ K_{3} = 1.8 $) }
\end{center} \label{fig.1:}
\end{figure}

For  $t >  \frac{2 }{3}\sqrt{K_{3}} \; \ln \beta$, the universe expands rapidly, %%@
eventually reaching an asymptotically de Sitter stage. Fig 1, gives the time variation %%@
of the Hubble parameter and its derivatives, for  $K_{3}=    1.8$ (which occurs if the %%@
particle number and species correspond to the minimal SU(5) model) and $\beta$ is given a %%@
value $10^{4}$.

3) It may be useful to check if the solution is stable under small linear perturbations. %%@
We write
\begin{equation}
H = H_{s} ( 1 + \delta)
\end{equation}
where $\delta$ is a small perturbation and $H_{s} = \frac{ \frac{2}{3} \alpha e^{\alpha %%@
t}}{(\beta + e^{\alpha t})}$ is the solution obtained above. Substituting this in %%@
equation (7) , we obtain the equation for $\delta$ :
\begin{equation}
\ddot{\delta} + A (t) \; \dot{\delta} + B(t) \; \delta = 0
\end{equation}
where 
\[
A(t) = \frac{ \alpha ( 1 + 2 e^{\alpha t})}{\beta + e^{\alpha t}} ,
\]
\begin{equation} 
 B(t) =  \frac{ 2 \alpha^{2} e^{\alpha t}}{(\beta + e^{\alpha t})^{2}}.
\end{equation}
Note that $\frac{\alpha}{\beta} < A(t) <  2 \alpha $  and $B(t) > 0$ for $ - %%@
\infty < t < \infty $.
It is difficult to solve the equation (12). However, if we assume that $\delta$ has a %%@
solution $\delta \sim e^{m t}$, we must have, for t negative,
\begin{equation}
m^{2}  + \frac{\alpha}{\beta} m \sim 0
\end{equation}
and $m $ cannot be positive, since $\alpha$ and $\beta$ are both positive. 
 Thus the solution seems to be %%@
stable under small perturbations at least in the negative t regime. 

4) The evolution can be described alternatively in terms %%@
of a scalar field  by substituting $ H = \phi^{2}$ in equation (7). This %%@
gives
\begin{equation}
\ddot{\phi} + 3 H \dot{\phi} + \frac{d V}{d \phi} = 0
\end{equation}
where $ V = - m^{2} \phi^{2} + \lambda \phi^{6}$ with $m^2 = \frac{1}{48 |K_{1}|}$ and $ %%@
\lambda = \frac{ K_{3}}{144 |K_{1}|}$. Note that $V(\phi)$ has the shape of a double %%@
well potential with a maximum at $\phi = 0$ and two minima for  $ \phi_{m} = \pm 
(\frac{m^{2}}{3 \lambda})^{1/4}$,  giving $V_{min} = - \frac{1}{ 72  |K_{1}| %%@
\sqrt{K_{3}}}$. Also note that $\phi = 0$ gives an unstable static universe while the %%@
evolution of the emergent universe is given by the path from $V(\phi) = 0 $   to %%@
$V_{min}$. However, $H$ here is time dependent and is determined self-consistently by the instantaneous value of $\phi$. The equation (15) does not make the problem simpler, although it looks very familiar and attractive. 

5) The prescribed evolution of the universe gets modified when particles are produced due to the expansion of the universe. We have considered above only massless, conformally coupled fields in conformally flat spacetimes, i.e., the conformally trivial situation where no particles are produced. The  perturbative calculations of particle production make use of small deviations from this conformal triviality. One considers either (i) a small mass and/or (ii) deviations from a conformally flat space time, say by assuming a Bianchi I spacetime deviating a little from a FRW spacetime or (iii) assuming a small non conformal coupling with $|\xi - \frac{1}{6}| = \epsilon$.   For simplicity, we shall consider here the  case of non conformal coupling. Let us consider a real massive scalar field satisfying the equation
\begin{equation}
\ddot{\phi} + 3 H \dot{\phi} +( m^{2} + \xi R ) \phi = 0
\end{equation}
with $|\xi - \frac{1}{6}| = \epsilon << 1$. As the scale factor  changes, the changing gravitational field feeds energy into the prturbed scalar field modes. As long as the mode frequency  of the field is greater than the Hubble expansion rate, a co-moving detector will not respond.  Modes with lower frequencies  will, however, be excited.   Thus the presence of mass makes the particle production process less efficient. To begin with we consider the case $ m = 0 $ and $ \epsilon \neq 0$.  
Zeldovich and Starobinsky, Starobinsky [3], Birrell and Davies [3] and Vilenkin [4] have studied this  case. 
The  rate of particle production per unit volume per unit conformal time $\eta$, $ \left( 
d \eta = \frac{dt}{a ( t )} \right)$, is given by 
\begin{equation}
\frac{dn}{d \eta} = \frac{1}{2} \epsilon^{2} R^{2}
\end{equation}
where $R$ is the scalar curvature.   This gives 
\begin{equation}
Y = \frac{2  a_{o}   }{{\epsilon}^{2}} \frac{dn}{dt} =
 \left[ \frac{36 \omega^{2} \alpha^{4} e^{2 \alpha t} (1 + 2 \omega e^{\alpha t})^{2}}{  (\beta + e^{ \alpha t})^{14/3}} \right]
\end{equation}
In fig. 2, we have plotted $ Y $  against $t$ for $\beta =  10,000$ and $K_{3} = 1.8 \; and \; 2.5$.  
We note that the rate of particle production peaks around $t \sim t_{o}$, as expected. Increase in the number of species of particles i.e., a higher value of $K_{3}$  will indicate a delay in particle production and also a wider peak. The dependence of the rate of particle production  on the parameter $\beta$ is shown in fig. 3. The absolute value of the peaks depends on the value of the scale factor $a(t)$ and the constant $a_{o}$ may be determined from the rate of particle production. Although we have considered zero mass particle we expect similar results even when $m$ is nonzero but small ( see Vilenkin [ 4 ]). During the subsequent evolution the particles produced thermalize and the universe enters eventually into a radiation dominated stage of the standard hot bigbang model.   The details of the process of particle production and the evolution of the universe will be considered elsewhere.

\input{epsf}
\begin{figure} 
\begin{center}
\vspace{6cm}
\special{eps: 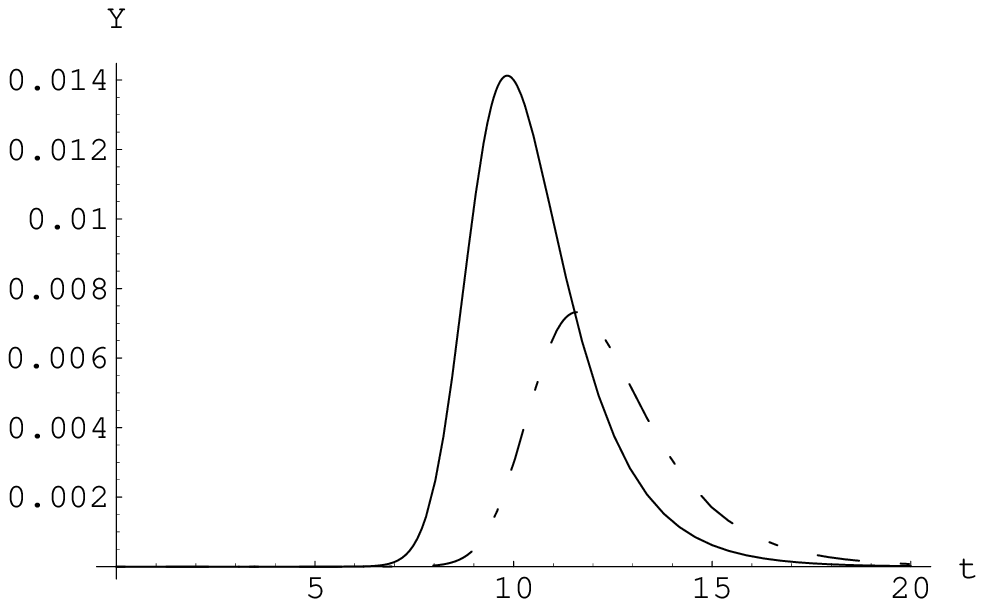 x=10cm y=6cm}
\caption{Dependence of rate of particle production on time for $\beta= 10000$ and $K_{3} = 2.5$ (broken line) and  $K_{3} = 1.8$ ( thin line ) }
\end{center} \label{fig.2:}
\end{figure}

\input{epsf}
\input{epsf}
\begin{figure} 
\begin{center}
\vspace{6cm}
\special{eps: 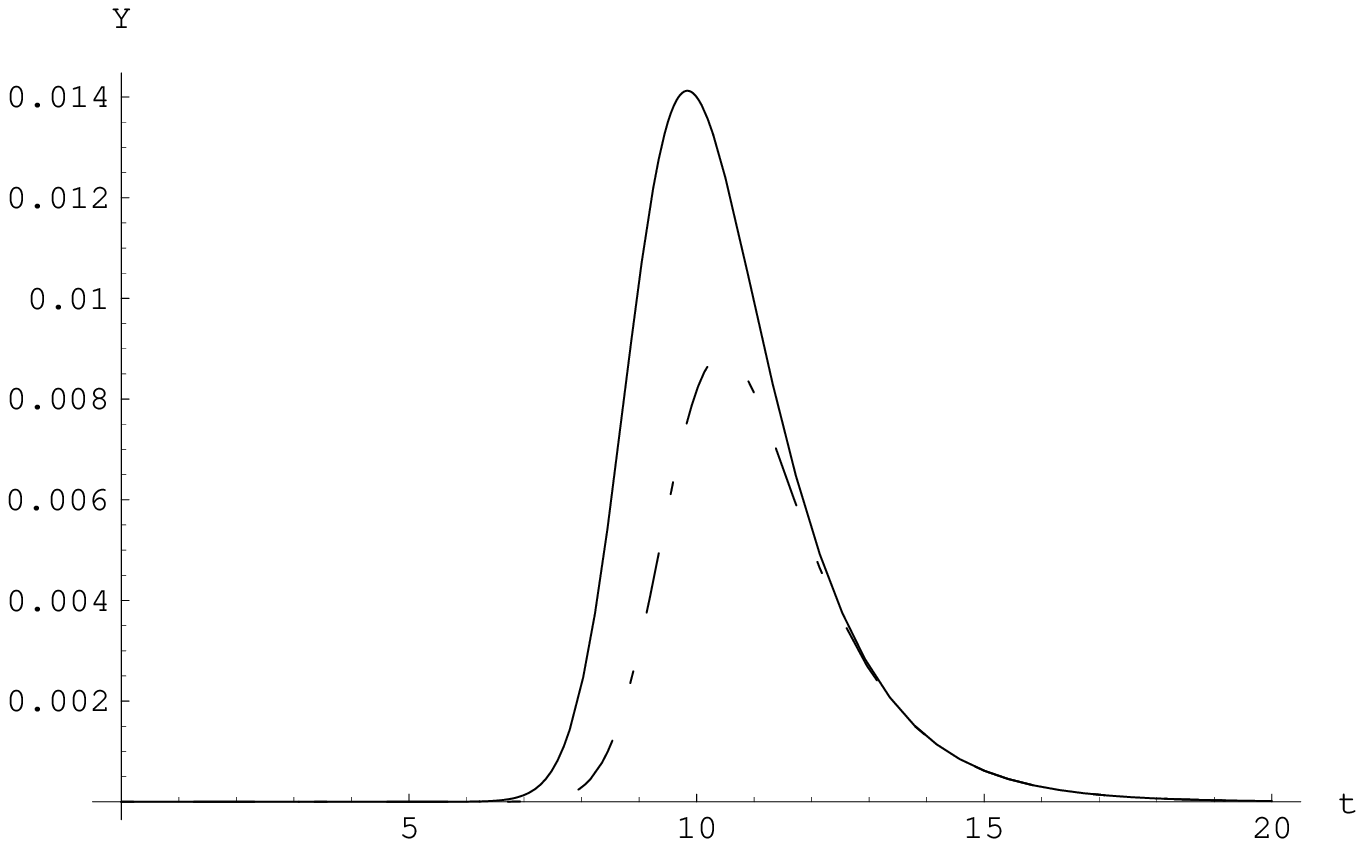 x=10cm y=6cm}
\caption{Dependence of rate of particle production on time for  $K_{3} = 1.8 $ and $\beta= 10000$ (thin line) and $\beta = 20,000$ ( broken line)   }
\end{center} \label{fig.3:}
\end{figure}

\section{DISCUSSION}

We have presented here a one parameter family of solutions of the Starobinsky model which 
describes an emergent universe.  The earlier work on Starobinsky model was done in the context of big bang singularity and the particle production was achieved by its oscillatory solution. The present solution on the other hand  describes the universe which is almost dormant during the infinite past period $ - \infty < t <  \frac{2}{3} \sqrt{K_{3}} \; \ln \beta$,  after which it undergoes a rapid expansion, see fig. 1.    The parameters $a_{o}$ and $\beta$ are related with initial conditions. They also introduce a new mass scale in the process of particle production. We have not considered here any specific theory of particle interactions which will determine the subsequent evolution of the universe as well as the details of large scale structure formation. 
 Whether the present  scenario can successfully explain the present observational data needs further study. However, the fact  
that one encounters solutions describing an emergent universe in different contexts may 
be a good reason for taking such solutions seriously, although the probabilities for such 
solutions may not be very high.

\vspace{0.5 cm.}

{\large \it  Acknowledgments :}

SM and BCP would like to thank the University of Zululand and the University of 
KwaZulu-Natal, South Africa for hospitality during their visit when a part of the work 
was done. They would also like to thank IUCAA, Pune  and IUCAA Reference Centre, North Bengal 
University for providing facilities. 

\pagebreak

\end{document}